\documentclass[doublecol]{epl2} 

\usepackage{amsmath}
\usepackage{amssymb}
\usepackage{pdfpages}

\usepackage[left=0.8in, right=0.6in, top=1in, bottom=1in]{geometry}

\title{Field Theory of Bayesian Rate Estimation}
\shorttitle{Field Theory of Bayesian Rate Estimation} 

\author{A. Auconi\inst{1}\thanks{andrea.auconi@gmail.com} \and A. Codello\inst{1,2} \and R. Burioni\inst{3} \and G. Caldarelli\inst{4,1,5}}
\shortauthor{A. Auconi \etal}

\institute{
  \inst{1} Ca' Foscari University of Venice, DSMN - via Torino 155, 30172 Mestre (Venice), Italy\\
  \inst{2} Instituto de F\'isica, Facultad de Ingenier\'ia, Universidad de la Rep\'ublica, 11000 Montevideo, Uruguay\\
  \inst{3} Dipartimento di Scienze Matematiche, Fisiche e Informatiche, Universit\`a degli Studi di Parma, Parco Area delle Scienze 7/A, 43124 Parma, Italy\\
  \inst{4} Consiglio Nazionale delle Ricerche, Institute of Complex Systems CNR-ISC, via dei Taurini 19, 00185 Rome, Italy\\
  \inst{5} London Institute of Mathematical Science, Royal Institution, 21 Albemarle St., London, UK
}

\abstract{
We address the statistical inference of a time-dependent rate of events in the framework of Bayesian field theory. 
This maps the problem to a Langevin equation which, beyond the local linear regime taken as reference, involves nonlinearities and an explicit dependence on the shape of the maximum \textit{a posteriori} estimator curve. 
We study the corresponding impacts in a perturbative expansion, {based on an analytically derived scaling relation for the order of shape corrections}.
We find that the pure nonlinearities dominate the mean and skewness.
Crucially, we uncover that the leading correction to the variance is driven by noise propagation from the signal's effective curvature.
We test the derived expansion and universal kernels against computationally expensive Monte Carlo simulations, and illustrate their applicability on real neural spike data.
}

\begin{document}

\maketitle

Reconstruction of signals from discrete noisy data is a ubiquitous challenge that spans neuroscience \cite{bialek1996field,mlynarski2021statistical,meshulam2025statistical}, biological sequences \cite{chen2021field,chen2024density}, bacterial chemotaxis \cite{vergassola2007infotaxis,auconi2022gradient,nakamura2024gradient}, cosmology \cite{ensslin2009information, ensslin2019information}, and more.
In these inverse problems, the setting is specified by a prior model of the unobserved signal and by a measurement model of the data likelihood conditional on the signal. The posterior distribution of the signal conditional on the measured data, that is, the signal reconstruction, is formally given by the Bayes formula.
The problem of how to actually compute, sample or approximate this posterior is the object of statistical inference \cite{petris2009dynamic,casella2024statistical}.

Bayesian Field Theory (BFT) \cite{bialek1996field,nemenman2002occam,ensslin2009information, ensslin2019information,kinney2014estimation,kinney2015unification} offers a principled framework for statistical inference by considering the posterior as a Gibbs-Boltzmann distribution, and leveraging field theory techniques like the perturbative expansion \cite{parisi1980perturbation,kardar2007statistical} to obtain approximations beyond the linear regime while gaining physical insight.

In the time-dependent rate estimation problem \cite{mora2019physical}, the prior model is typically a geometric Brownian motion (GBM) \cite{karatzas2014brownian}, while the measurement model is {by definition} a Poisson process \cite{risken1996fokker}. 
A similar structure is obtained in the 1D probability density estimation \cite{bialek1996field,nemenman2002occam,kinney2014estimation,kinney2015unification} differing by a normalization constraint.
The volatility of GBM is an hyperparameter which, {assuming it to be a constant,} it can be reliably estimated over a sufficiently long measurement time. 
Approximation methods for the opposite data-poor scenario have been studied in \cite{chen2018density}.

For a fixed known volatility, the posterior rate path distribution can be exactly mapped to the equilibrium Langevin field fluctuations in a potential \cite{parisi1980perturbation,kardar2007statistical}.
The configuration that minimizes the potential corresponds to the {maximum \textit{a posteriori} estimator (MAP)} \cite{petris2009dynamic,casella2024statistical} for the log-process, while the fluctuations around it encode the distribution \cite{bialek1996field}.
In principle, a full characterization of such distribution or any of its observables can be performed by computationally expensive simulations of the discretized Langevin equation, {analogous to Markov Chain Monte Carlo (MCMC) methods \cite{petris2009dynamic,casella2024statistical}.}



In this paper, we {study} the perturbative expansion of the Langevin equation {corresponding to} the rate estimation problem.
After reviewing the local linear regime, we derive the first two perturbative correction orders, {which impact} the mean, variance, and skewness.
We explicitly decompose them into pure nonlinearities arising solely from the non-harmonic local potential, and shape corrections arising from the MAP curve geometry.
{We establish the relative order of these corrections based on a scaling relation between Bayesian uncertainty and MAP curve fluctuations, that we here derive considering the limiting Kalman smoother dynamics.}

{In the perturbative regime,} the derived universal kernels governing such shape corrections allow to immediately improve the posterior estimation 
{compared to the local linear approximation.} 
Furthermore, they reveal a key property of the uncertainty dynamics, 
i.e., that the effective MAP curvature impacts the posterior uncertainty by noise propagation from neighboring regions,  expanding the error tube at peaks of the MAP curve relative to the local linear regime.



\section{The rate estimation problem}
Consider observing a spike train $D\equiv (t_1, t_2, ..., t_m)$ of $m$ events over a time interval $[0, T]$. The standard description \cite{mora2019physical,vergassola2007infotaxis,auconi2022gradient,petris2009dynamic,casella2024statistical} assumes the existence of an underlying time-dependent rate $r_t$ from which such events were stochastically generated according to an inhomogeneous Poisson process, meaning that at each time interval of infinitesimal length $dt$ a single event may occur with probability $r_t dt$. The dynamics of this rate is modeled by a GBM, that is the simplest continuous stochastic process satisfying the positivity constraint $r_t >0$. It is defined by the stochastic differential equation $dr = \sigma r dW$ written in the Ito interpretation \cite{karatzas2014brownian}, where $dW$ are standard Brownian motion increments and $\sigma$ is the volatility parameter. 
It is convenient to study its logarithm $s \equiv \ln r$ which follows $ds = \sigma dW -\sigma^2 dt/2$ from Ito's Lemma. 

The prior probability density of paths $\boldsymbol{s} \equiv \left\{ s_{j dt}  \right\}_{j=0}^{T/dt}$ in the vanishing time discretization limit $dt\rightarrow 0$ can be written in exponential form as, up to a constant,
\begin{equation}\label{prior}
    - \ln p(\boldsymbol{s}) = \frac{s_T-s_0}{2} +  \frac{1}{2\sigma^2 dt} \int_0^T ds \,ds,
\end{equation}
where the first term accounts for the drift in log-scale required for GBM to be a martingale, while the second term accounts for its fluctuations through the quadratic variation $\int_0^T ds\, ds \equiv \sum_{j=1}^{T/dt} (s_{j dt}-s_{(j-1)dt})^2$ according to Lévy's characterization theorem \cite{karatzas2014brownian}. The prior on the initial condition at $t=0$ is taken flat and omitted, $p(s_0) \sim \exp [-b (s_0-a)^2]$ with $b\rightarrow 0^+$.

The observed spike train $D$ provides information on the path $\boldsymbol{s}$ based on the Poisson measurement model, whose log-likelihood is written as, up to a constant,
\begin{equation}\label{measurement model}
     \ln p(D \vert \boldsymbol{s}) = \sum_{t\in D} s_t -\int_0^T dt\, e^{s_t} ,
\end{equation}
where the first term accounts for the observed events $D$, while the second term accounts for the absence of other events.
From Bayes formula, the log-posterior distribution is then, up to a constant,
\begin{equation}\label{posterior}
    \ln p(\boldsymbol{s} \vert D) = \ln p(\boldsymbol{s}) +\ln p(D \vert \boldsymbol{s}) \equiv -V,
\end{equation}
which is formally a Gibbs-Boltzmann distribution in terms of the potential $V\equiv V[\boldsymbol{s}](D)$. 
Its functional derivative is 
\begin{equation}\label{functional derivative}
        -\frac{\delta V}{\delta s_t} = \frac{1}{\sigma^2} d^2_t s_t
        +\frac{1}{2}(\delta_{t,0} -\delta_{t,T}) + \sum_{w\in D} \delta_{t,w}  -e^{s_t},
\end{equation}
where the Laplacian $d^2_t s_t$ is defined by the Neumann boundary conditions $s_{-dt}=s_0$ and $s_{T+dt}=s_T$, and $\delta_{t,t'}\equiv \delta (t-t')$ denotes the Dirac delta.




\section{Classical solution}
\begin{figure}
    \centering    \includegraphics[width=0.95\linewidth]{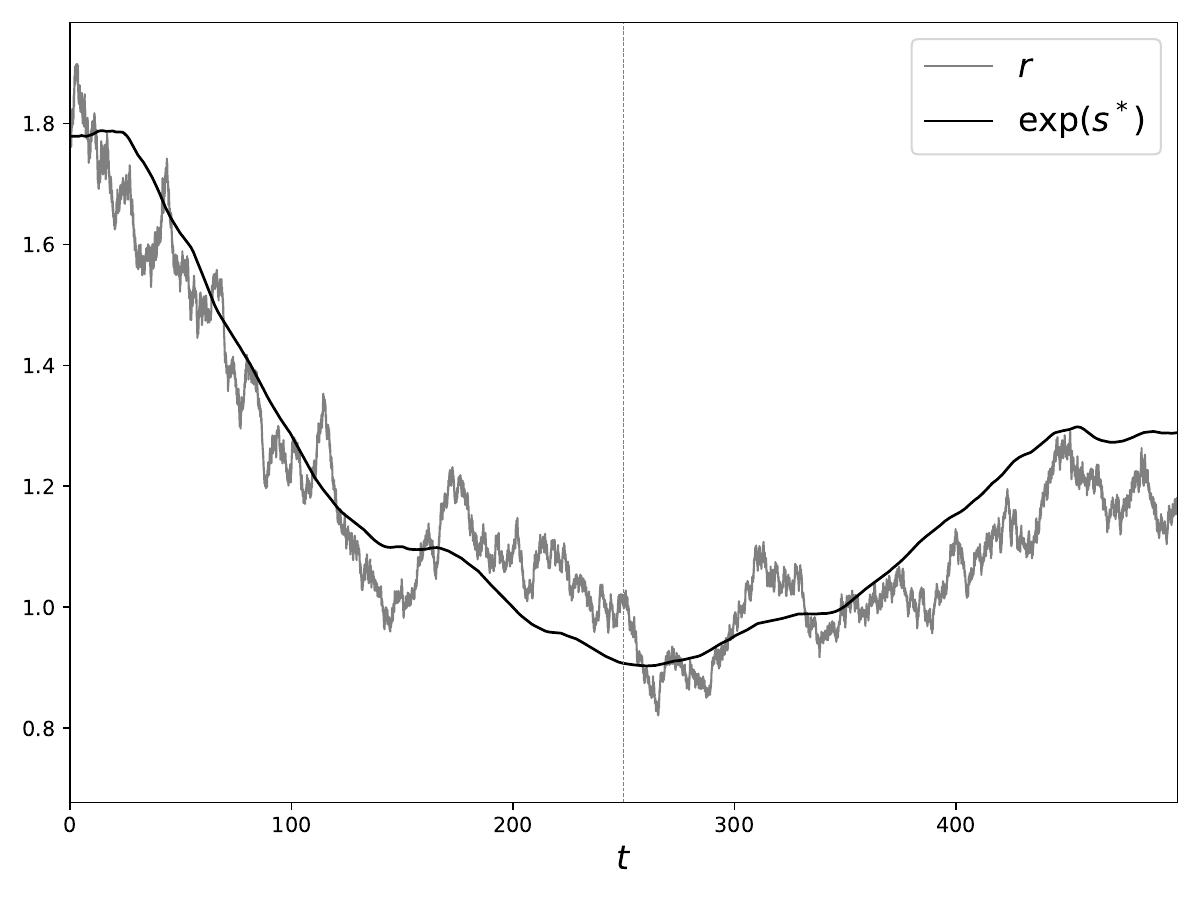}
    \caption{The time-dependent rate $r_t$ (gray line) generates a spike train $D$, {based on which} the posterior probability $p(\boldsymbol{s} \vert D)$ of the log-process $s_t\equiv \ln (r_t)$ {is defined.}  The MAP curve $\exp (s^*_t)$ (black line) {is the corresponding optimal path,} found by numerically solving Eq. \eqref{Classical}. The volatility parameter is $\sigma = 0.02$, {and we fix the underlying rate to satisfy $r_0 = 1$, so that the local linear regime at $t=0$ is characterized by estimated rate $\alpha\approx r_0$, variance $\nu = \sigma/(2\sqrt{\alpha}) \approx 0.01$, and timescale $\tau=1/(\sigma\sqrt{\alpha}) \approx 50$. The events $u\in D$ correspond to the corner points of $s^*_t$, not visible at this resolution because their number within a timescale diverges as $\alpha\tau=1/(2\nu)$ in the perturbative limit $\nu\rightarrow 0$, see discussion after Eq. \eqref{Gauss covariance}}.}
    \label{fig:MAP}
\end{figure}

The MAP curve $\boldsymbol{s}^*$ is defined by
\begin{equation}\label{Classical}
       \frac{\delta V}{\delta s_t} \bigg|_{\boldsymbol{s}^*} = 0,
\end{equation}
that is a second order differential equation which can be solved numerically, see Fig. (\ref{fig:MAP}), {provided that at least one spike was observed, $m\geq 1$.}
Indeed, time-integrating once this equation yields the compatibility condition
$m = \int dt \, \exp(s^*_t)$,
which excludes a finite solution for $m=0$. While this issue could be corrected by specifying an informative prior $b>0$, we will instead consider only the case of a non-empty dataset $m\geq 1$.
Since the potential $V$ is convex, the MAP curve $\boldsymbol{s}^*$ is also unique in this case.

\section{Langevin fluctuations}
Let us denote by $x_t\equiv s_t-s^*_t$ the displacement of a path from the MAP curve.
The posterior distribution $p(\boldsymbol{s} \vert D)\sim \exp(-V)$ is sampled by the corresponding Langevin equation \cite{parisi1980perturbation,kardar2007statistical},
\begin{equation}\label{SPDE}
    \partial_u x_{t, u} -\eta_{t, u} = - \frac{\delta V}{\delta x_{t, u}} \, = \frac{1}{\sigma^2} \partial^2_t x_{t, u} -e^{s^*_t} (e^{x_{t, u}} -1) . \,
\end{equation}
This stochastic partial differential equation (SPDE) \cite{hairer2009introduction,kardar2007statistical,lord2014introduction} describes the evolution of the field $x_{t, u}$ in fictitious time $u$, where the signal time $t$ plays the role of the spatial coordinate.
The field is driven by space-time Gaussian white noise $\eta_{t, u}$ with mean $ \langle \eta_{t, u} \rangle =0$ and covariance $ \langle \eta_{t, u} \eta_{t', u'} \rangle = 2 \delta_{t,t'} \delta_{u,u'}$.
The drift term involves an exponential nonlinearity $e^{x_{t, u}}$, and a space-dependent coefficient $e^{s^*_t}$.

Let us consider any particular time point $t_e \in [0,T]$ enough distanced from the borders, for which we wish to estimate the marginal posterior $p(x_{t_e}\mid D)$. For ease of notation, let us shift the time axis to have $t_e = 0$, and simply denote the local moments as $\langle x^k \rangle \equiv \langle x_0^k \mid D \rangle$. 




\section{The local linear regime}

The {homogeneous} Gaussian approximation $x_{t,u} \approx y_{t,u}$ to the general Eq. \eqref{SPDE} is obtained by linearizing the drift $\exp(x_{t, u}) -1 \approx x_{t,u}$, and taking a constant coefficient $\exp(s^*_t)\approx \exp(s^*_0) \equiv \alpha$. This dynamics is solved by the convolution
\begin{equation}\label{convolution}
    y_{t, u} = G_{t-t', u-u'} \, \eta^{t', u'},
\end{equation}
where the integration convention is used for variables repeated in superscript and subscript, $a_x b^x\equiv \int_{-\infty}^{+\infty} dx\, a_x b_x$. 
The Green function $G_{t, u}$, see, e.g.,  \cite{kardar2007statistical,hairer2009introduction}, is reviewed in Appendix \ref{app:Gauss} for the reader's convenience.
The stationary covariance $C_{t,u}\equiv \langle y_{h,q}y_{h+t, q+u} \rangle$ evaluated at $u=0$ gives
\begin{equation}\label{Gauss covariance}
   C_{t, 0} = \nu \, e^{-|t|/\tau} ,
\end{equation}
with variance $\nu \equiv C_{0,0} = \langle y^2 \rangle = \sigma/(2\sqrt{\alpha})$, and correlation timescale $\tau\equiv 1/(\sigma \sqrt{\alpha})$.

The local linear regime is a good approximation when the rate of events is much larger than its squared volatility, $\alpha \gg \sigma^2$. Indeed, in this limit the fluctuations are small, $\nu \ll 1$, and this makes the linear approximation $e^x-1\approx x$ become accurate.
Moreover, as observed in \cite{mora2019physical}, in this limit the correlation timescale is much smaller than the characteristic timescale of the underlying rate diffusion, $\tau \ll \sigma^{-2}$, and this makes the approximation of using the local coefficient $e^{s_t}\approx\alpha$ become accurate. The differing factor $1/2$ in the local variance $\nu$ compared to Ref. \cite{mora2019physical} is expected as we are conditioning on the full time series and not only on past events.
Note also that in this limit the correlation timescale is much larger than the inverse rate, $\tau \gg \alpha^{-1}$, meaning that each correlated region is made of a large number of events. 


\section{The perturbative expansion}
Let us define the deviation from the reference mean-reverting coefficient $\alpha\equiv \exp[s^*_0]$ as $f_{t} \equiv \exp[s^*_t]-\alpha$.
The perturbative expansion starts from the formal implicit integration with respect to the local Green function of those terms in Eq. \eqref{SPDE} that were neglected in the local linearized dynamics,
\begin{multline}\label{Perturbative}
    x_{t, u} = y_{t, u} - G_{t-t', u-u'} \alpha \left[ \exp(x^{t',u'}) -1^{t',u'} -x^{t',u'} \right] \\ - G_{t-t', u-u'} f^{t'} \left[ \exp(x^{t',u'}) -1^{t',u'} \right].
\end{multline}
We then Taylor-expand the nonlinearities and iteratively substitute the $x$ variables on the right-hand side by using Eq. \eqref{Perturbative} itself. Such recursive process is repeated until the field is expressed solely in terms of Gaussian products $y_{0,0} y_{t,u}y_{t',u'}...$ up to the desired order.
When computing moments, the order is specified with respect to the small parameter $\nu \equiv \langle y^2 \rangle {=\sigma/(2\sqrt{\alpha})}$, to which we also relate the MAP shape corrections from $f_t\neq 0$ according to a scaling {relation} as detailed below.


\section{Homogeneous case}

Let us first consider the case $f_t=0$, that means having a constant mean-reverting coefficient $\alpha$ which makes expectations independent of the fictitious space $t$. Let us denote the corresponding process as $\widetilde{x}$.
Taking the expectation of Eq. \eqref{Perturbative} considering the integral $G_{t-t', u-u'} 1^{t',u'}=1/\alpha$ yields the fluctuation theorem
$\langle \exp(\widetilde{x}) \rangle = 1$, a form that is widespread in stochastic thermodynamics \cite{seifert2008stochastic,sagawa2012nonequilibrium,ito2013information,auconi2019information}, and which serves here as a constraint between moments in the perturbative expansion.
In particular, by expanding the exponential to second order we recover the leading order correction to the mean $\langle \widetilde{x}\rangle \approx -\nu /2$, that is driven by the asymmetry of the drift $1-\exp(\widetilde{x})$ which favors downward fluctuations, and it is understood as the estimation of lower rates is slower. An analogous correction term was found in Ref. \cite{bialek1996field} for the related problem of the normalized 1D probability density estimation.

Let us now consider terms up to order $\mathcal{O}(\nu^2)$ .
For the perturbative expansion of the variance we find a pure nonlinearity correction of
\begin{equation}\label{nonlinearities Var corr}
   \langle \widetilde{x}^2 \rangle -\langle \widetilde{x} \rangle^2 -  \nu 
    = \frac{1}{9} \nu^2   ,
\end{equation}
where Wick's theorem was used to express the higher order moments of the Gaussian variables as covariances, and the resulting integrals evaluated with Wolfram Mathematica \cite{wolfram2003mathematica}. See Appendix \ref{app:diagrams_homogeneous} for a diagrammatic illustration of the perturbative expansion, and the Supplementary Materials (SM) for more details on the derivations, which includes references to \cite{barbieri2011trajectories,hohenberg1977theory,gradshteyn2014table,auconi2024interaction}.
The pure nonlinearity correction is positive, meaning more uncertainty, driven by the sub-linear drift for downward fluctuations and by the mean $\langle \widetilde{x} \rangle < 0$ shifted away from the exponential wall for upward fluctuations.

For the third and fourth moment we find respectively $\langle \widetilde{x}^3 \rangle = -\frac{11}{6} \nu^2$, and $\langle \widetilde{x}^4 \rangle = \langle y^4 \rangle = 3 \nu^2$, while higher moments are negligible at this order. The expectation is then computed by expanding the fluctuation theorem to fourth order, and it gives $\langle \widetilde{x}\rangle = -\nu /2$.



\section{{Scaling relation}}

In the limit $\nu\rightarrow 0$, the variation of the log-rate $s_t$ over the correlation timescale $\tau$ is statistically small, $s_\tau -s_0 = \mathcal{O}(\sqrt{\nu})$, at the same order of the estimation errors $s_0^*-s_0 = \mathcal{O}(\sqrt{\nu})$ and $s_\tau^*-s_\tau = \mathcal{O}(\sqrt{\nu} )$. This implies that also the MAP variation is small on the correlation timescale, {at most of that same} order, $s^*_\tau-s^*_0= \mathcal{O}(\sqrt{\nu})$.
Then we can linearize the MAP dynamics Eq. \eqref{Classical}, see the SM, to obtain the \textit{local} 
Kalman smoother (or Wiener filter) \cite{petris2009dynamic,casella2024statistical},
\begin{equation}\label{Kalman smoother}
    \exp(s^*_t) \approx h_t \equiv \frac{1}{2\tau} \sum_{w\in D} \exp\left(-\frac{|t-w|}{\tau}\right),
\end{equation}
{based on the local timescale $\tau = [\sigma^2 \exp(s^*_0)]^{-1/2}$. Note that this approximation becomes accurate on any large region $t \sim c\tau$ with fixed finite $c$ in the limit $\nu\rightarrow 0$.}
Taking expectations $\mathbb{E}$ over the joint rate dynamics and events' realization conditional on a fixed $r_0$, to leading order we find the relation, {see the SM,
\begin{equation}\label{Eq12}
\mathbb{E}\left[f_t^2 \mid r_0\right] \approx 2 \alpha^2 \nu  \left(\frac{|t|}{\tau} -1 +e^{-|t|/\tau}\right) \Big|_{\alpha=r_0},
\end{equation}
where the parameters $\alpha$, $\nu$, and $\tau$ on the right-hand side are evaluated at the condition $\alpha = r_0$.
This relation captures the transition from a short-time ballistic regime $f_t \sim t$ for $t\ll\tau$ to a long-time diffusive regime $f_t \sim \sqrt{t}$ for $t\gg\tau$. Clearly, at fixed $\nu$ the approximation eventually breaks down for even longer times $t\rightarrow \infty$. Crucially, evaluated around the Bayesian timescale $t\sim\tau$, Eq. \eqref{Eq12} establishes the exact asymptotic scaling}
\begin{equation}\label{scaling}
    f_\tau/\alpha = \mathcal{O}(\sqrt{\nu}).
\end{equation}
{Given that convolutions in the perturbative expansion are characterized by the timescale $\tau$, this scaling Eq. \eqref{scaling} tells us that factors $f_t/\alpha$ have to be treated at the same order as the Gaussian standard deviation $\sqrt{\nu}$.}

\begin{figure}
    \centering    \includegraphics[width=0.95\linewidth]{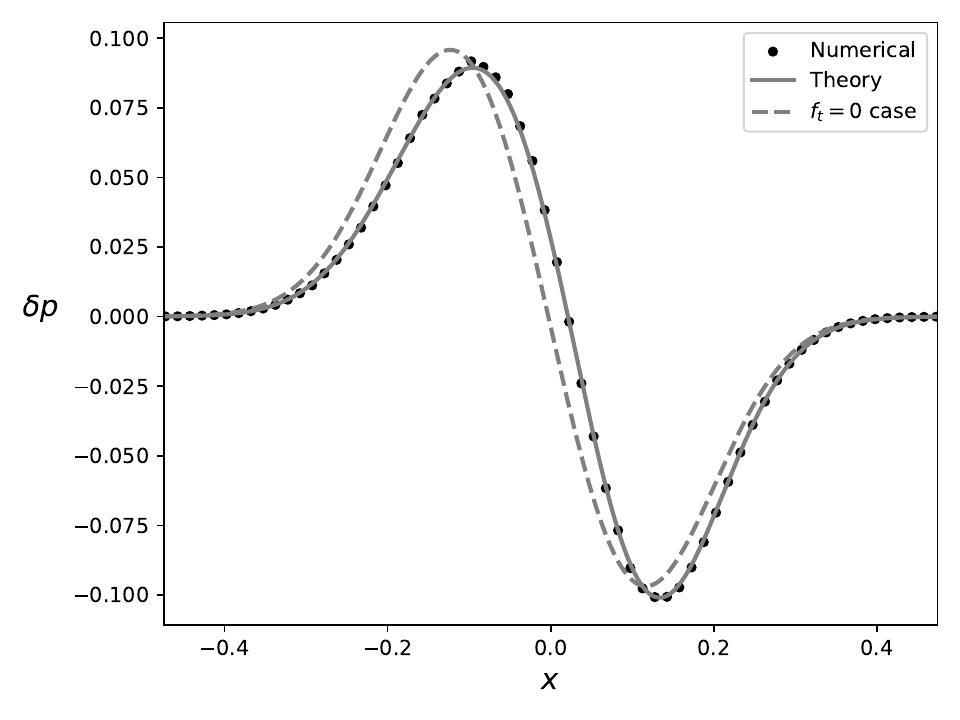}
    \caption{The deviation $\delta p$ from the local linear regime is {sampled} numerically (dots) from simulations of the SPDE Eq. \eqref{SPDE} with a Laplacian-implicit Euler method, and compared to the Gram-Charlier expansion corresponding to the perturbative expansion moments from Eqs. \eqref{Variance impact}-\eqref{path impact on expectation} (solid line). The impact of shape corrections is highlighted by comparison to the homogeneous case  $f_t=0$ (dashed line). The parameters and MAP curve $f_t$ are those of Fig. (\ref{fig:MAP}). More examples in the SM.}
    \label{fig:numerical}
\end{figure}

\section{Shape corrections}
Equipped with this scaling {relation}, we now consider the impacts of the MAP curve deviation $f_t$ on the local posterior $p(x)\equiv p(x_{0}|D)$.
Let us first note that a space-dependent coefficient in the SPDE Eq. \eqref{SPDE} makes the marginals space-dependent inducing a violation of the homogeneous fluctuation theorem, $\langle e^x \rangle \neq 1$, which decouples the moments at least in this simple form.

The perturbative expansion of the variance yields
\begin{equation}\label{Variance impact}
   \langle x^2 \rangle -\langle x \rangle^2 -\nu =  \frac{1}{9} \nu^2  + J_t f^t  + K_{t, t'} f^t f^{t'} , 
\end{equation}
where the $J_t$ and $K_{t, t'}$ kernels' definitions and {corresponding Feynman diagrams} are provided in Appendix \ref{app:diagrams_shape}.
The leading shape correction term $J_t f^t$ accounts for the larger (smaller) noise propagating from neighboring regions with lower $f_t<0$ (higher $f_t>0$) damping coefficient.
Its perturbative order according to the scaling {relation Eq. \eqref{scaling}} is $J_t f^t = \mathcal{O}\left(\nu^{3/2}\right)$, which statistically dominates the nonlinearity correction $\nu^2/9$.


Physically, the shape correction $J_tf^t$ modulates the posterior variance by noise propagation, thereby counteracting the local extremes to dampen the spatial fluctuations of the error envelope predicted by the local linear theory.
In fact, this term can be reproduced to the required order {maintaining the form} $\sigma/(2\sqrt{\widetilde{\alpha}}) = \nu +  J_t f^t$ {but} in terms of an effective damping
\begin{equation}\label{effective damping}
    \widetilde{\alpha} \equiv -(\nu^2 \tau)^{-1} J^t \exp(s^*_t),
\end{equation}
where we used $J_t 1^t=-\nu^2 \tau$.
This means to integrate the MAP curve $\exp(s^*_t)$ with respect to the kernel $-(\nu^2 \tau)^{-1}J_t$, which acts over short-medium timescales bounded by the correlation timescale $\tau$.
Given the symmetry $J_t=J_{-t}$, and that by definition $f_0=0$, the first polynomial contribution in $J_t f^t$ comes from the quadratic term $f_t\sim t^2$, giving the interpretation of an effective curvature.


Consider the exponential smoothing kernel $G_{t, u} 1^u = \nu \exp(-|t|/\tau)$.
Let us define the smoothed version of any function of time $z_t$ as $\overline{z_t} \equiv G^{-t', -u'} 1_{u'} z_{t+t'}$, and similarly for time-lagged products $\overline{z_t z_{t'}} \equiv G^{-t'', -u''} 1_{u''} z_{t+t''} z_{t'+t''} $.
The shape correction to the expectation is
\begin{multline}\label{path impact on expectation}
    \langle x \rangle -  \langle \widetilde{x} \rangle 
   \\= - \frac{\alpha}{2} \left[ J_{t} \,\overline{f^{t}} +K_{t, t'}\, \overline{f^{t} f^{t'}}
    - J_{t} \, \overline{f_0 \left(\overline{f^{t}} -f^{t}/\alpha \right)}   \right] ,
\end{multline}
see Appendix \ref{app:diagrams_shape} for the corresponding Feynman diagrams.
The skewness is not impacted at this order.

The derived expansion Eqs. \eqref{Variance impact}-\eqref{path impact on expectation} have been tested by comparing the corresponding Gram-Charlier expansion to the numerical distribution from simulations of the full SPDE Eq. \eqref{SPDE}, see Fig. (\ref{fig:numerical}). 



\begin{figure}
    \centering    \includegraphics[width=0.95\linewidth]{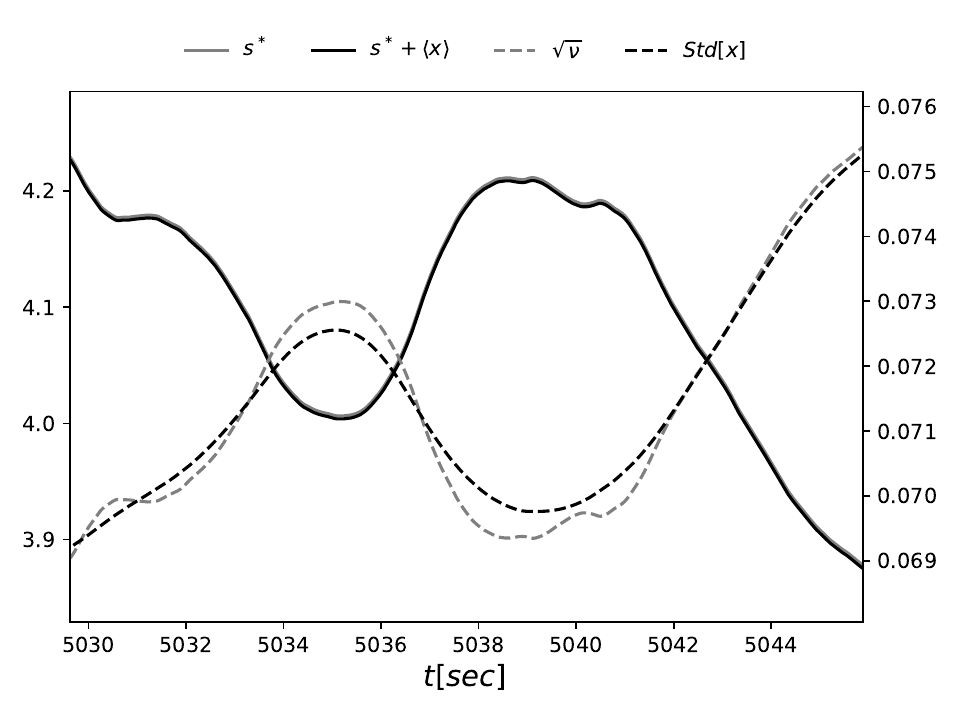}
    \caption{The MAP log-rate $s^*_t$ and expected log-rate $s^*_t+\langle x\rangle$ for the most active neuron in the dataset \cite{siegle2021survey} is shown on the left axis, with the volatility fixed to its optimal value $\sigma^* \approx 0.08 \,\text{sec}^{-1/2}$. The posterior standard deviation $\text{Std}[x]\equiv (\langle x^2\rangle -\langle x \rangle^2)^{1/2}$, shown on the right axis, expands significantly compared to the local linear solution $\sqrt{\nu}$ at peaks of the MAP curve, see Eq. \eqref{effective damping}. More neuron data examples in the SM.
    }
    \label{fig:neuron}
\end{figure}

\section{Illustration on neural spike data}

We include a purely illustrative application of the derived expansion on real neural spike data obtained from the Allen Brain Observatory \cite{siegle2021survey}. We selected a highly active and stable neuron from the dataset to ensure a robust estimation. We estimated the volatility hyperparameter posterior distribution $p(\sigma |D)$ with the Laplace approximation \cite{bialek1996field,kinney2015unification,shun1995laplace}, and it results to be sufficiently peaked around its maximum, see the SM, allowing us to fix $\sigma\approx 0.08 \,\text{sec}^{-1/2}$. The reconstructed rate uncertainty $\langle x^2 \rangle -\langle x \rangle^2$ illustrates the noise propagation mechanism of the leading shape correction $J_t f^t$, as the posterior variance expands significantly at peaks relative to the local linear approximation, effectively dampening its temporal fluctuations.

{
\section{Beyond the GBM Eq. \eqref{prior} prior}
Gaussian processes with time-dependent coefficients introduce terms into Eq. \eqref{Perturbative} that are linear in $x_{t,u}$, inducing corrections to the posterior variance of order $\mathcal{O}(\nu^2)$, see the SM, where we also show that smooth nonlinear drifts in the prior enter only at $\mathcal{O}(\nu^3)$.
While the $\mathcal{O}(\nu^2)$ terms in Eq. \eqref{Variance impact} are thus model-dependent, this establishes the leading shape correction $J_tf^t = \mathcal{O}(\nu^{3/2})$ as a general feature of non-parametric rate estimation across this Gaussian universality class.}
Priors outside this class, such as processes with jumps or anomalous diffusion, correspond to different fixed points and are not covered by the present expansion. We note that the Poisson likelihood, by contrast, is not a modeling assumption but the definition of the rate estimation problem itself.

\section{Discussion}


We analyzed the rate estimation problem in the Bayesian field theory framework {pioneered in Ref. \cite{bialek1996field}}, and derived a {second order} perturbative expansion of the corresponding Langevin equation {in the small fluctuations limit}. We {uncovered the role of} shape corrections, i.e., those arising from the MAP curve geometry, and found that they statistically dominate the variance corrections based on a scaling {result analytically obtained from the limiting} Kalman smoother's statistics. 
{To leading order, they are interpreted as} a noise propagation mechanism, driven by the signal's effective curvature, that dampens the spatial fluctuations of the variance envelope predicted by the local linear theory. 
In fact, the corresponding universal kernel generalizes the concept of curvature correction, previously identified in the information theory of chemotaxis \cite{auconi2022gradient}, {to arbitrary shapes}.

{To the best of our knowledge, a perturbative expansion of an SPDE with space-dependent coefficients whose variation statistically depends on the perturbation parameter itself, that is the mathematical structure of rate estimation we uncover with the scaling relation Eq. \eqref{scaling}, was not present in the literature.}

The formal similarity between our SPDE Eq. \eqref{SPDE} and the stochastic cable equations for single neurons \cite{gowers2023upcrossing, gowers2025resonances} suggests that our approach could be extended to perturbatively describe dendrites with space-dependent synaptic conductance.

While exact inference typically requires computationally expensive sampling of the posterior via MCMC methods \cite{petris2009dynamic,casella2024statistical}, 
{if the data is in the small fluctuations limit $\nu\ll 1$}
our analytical expansion yields an immediate estimate based on the derived universal kernels.  
Our framework, however, like all perturbative expansions, is accurate only in the vicinity of the reference local linear regime. 
Looking forward, this perturbative approach could be extended to the level of hyperparameter estimation \cite{chen2021field}, providing a path toward explainable higher-order analytical refinements.

\makeatletter
\newcommand{\eplappendix}[3]{%
  \section{Appendix #1: #2}%
  \setcounter{equation}{0}%
  \renewcommand{\theequation}{#1.\arabic{equation}}%
  \def\@currentlabel{#1}\label{#3}}
\makeatother

\eplappendix{A}{Gaussian propagator and correlator}{app:Gauss}

The local linearized dynamics is {the massive stochastic heat equation \cite{hairer2009introduction,kardar2007statistical,lord2014introduction}} written
\begin{equation}\label{Basic SPDE}
    \partial_u y_{t, u} = \frac{1}{\sigma^2} \partial^2_t y_{t, u} -\alpha y_{t, u} + \eta_{t, u},
\end{equation}
and its solution $y_{t, u} = G_{t-t', u-u'} \, \eta^{t', u'}$ involves the Green function
\begin{equation}\label{G}
    G_{t, u} = \frac{\sigma \mathbb{I}_{u\geq 0}}{2\sqrt{\pi u}} \exp \left( -\alpha u -\frac{\sigma^2 t^2}{4 u} \right), 
\end{equation}
where $\mathbb{I}_{u\geq 0}$ is the Heaviside step function ensuring causality.
The stationary covariance is computed as
\begin{multline}\label{C}
    C_{t,u} = 2G^{-t', -u'} G_{t-t', u-u'} \\
    = \frac{\sigma}{2\pi} \int dk \, \frac{\cos(k \sigma t) }{\alpha +k^2} e^{ -(\alpha+k^2) \,\vert u \vert }.
\end{multline}


\eplappendix{B}{Diagrams in the homogeneous case}{app:diagrams_homogeneous}

To illustrate the perturbative expansions, which have also been explicitly carried out in the SM, we consider Feynman diagrams \cite{kardar2007statistical,parisi1980perturbation,tauber2014critical} where the observation point $(0,0)$ is denoted by $\circ$, and the $n$-th order nonlinearities in Eq. \eqref{Perturbative} are represented by integration vertices $\bullet$ carrying a factor of $-\alpha/ n!$. These are connected via directed $G$ (arrows) and undirected $C$ (plain lines) propagators, corresponding respectively to Eq. \eqref{G} and Eq. \eqref{C}.

According to these rules, the pure nonlinearity correction to the variance is, up to $\mathcal{O}\left(\nu^2\right)$,
\begin{multline}\label{Feynman Var homogeneous}
\left\langle \widetilde{x}^2 \right\rangle -\langle \widetilde{x} \rangle^2 - \nu = \frac{\alpha^2}{2} G^{-t, -u} C_{t-t', u-u'} \\
\times \left( G^{-t', -u'} C_{t-t', u-u'} 
  + 4 G^{t-t', u-u'}  C_{t', u'} \right) \quad \\= \quad 
\vcenter{\hbox{\includegraphics[page=1]{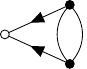}}}
\quad + \quad
\vcenter{\hbox{\includegraphics[page=2]{Figures.pdf}}}
\quad 
  = \quad \frac{1}{9} \nu^2,
\end{multline}
where the derived integrals have been evaluated with Wolfram Mathematica \cite{wolfram2003mathematica}, see the SM.
For the skewness we find, up to $\mathcal{O}\left(\nu^2\right)$,
\begin{multline}\label{Feynman Skew}
\langle \widetilde{x}^3 \rangle - 3\langle \widetilde{x}^2 \rangle \langle \widetilde{x} \rangle + 2\langle \widetilde{x} \rangle^3 = - 3\alpha \, G^{-t,-u} C_{t,u}^2 
\\\quad = \quad
\vcenter{\hbox{\includegraphics[page=3]{Figures.pdf}}}\quad
  = \quad  -\frac{1}{3} \nu^2 .
\end{multline}

\eplappendix{C}{Diagrams for the shape corrections}{app:diagrams_shape}

The impact of the MAP curve $f_t$ in the expansion is represented by an external source vertex $\star$ of factor $-f_t/n!$. 
The corresponding shape corrections to the variance read, up to $\mathcal{O}\left(\nu^2\right)$ terms,
\begin{multline}\label{Feynman Var shape}
 \left\langle x^2 \right\rangle -\langle x \rangle^2 - \left\langle \widetilde{x}^2 \right\rangle + \langle \widetilde{x} \rangle^2    =   -2 G_{-t, -u} {C_{t,}}^u f^t 
\\ + G_{-t, -u} f^t f^{t'}  \left( 2 {G_{t-t',}}^{u-u'} C_{t', u'}
+ G_{-t', -u'} {C_{t-t',}}^{u-u'} \right) \\
= \quad 
\vcenter{\hbox{\includegraphics[page=4]{Figures.pdf}}}
\quad + \quad
\vcenter{\hbox{\includegraphics[page=5]{Figures.pdf}}}
\quad + \quad
\vcenter{\hbox{\includegraphics[page=6]{Figures.pdf}}}
\\ \equiv J_t f^t  + K_{t, t'} f^t f^{t'}.
\end{multline}
We can integrate with respect to the fictitious time $u$ and write explicitly the leading shape kernel $J_t$, that is the leftmost graph in Eq. \eqref{Feynman Var shape}, as
\begin{equation}\label{J_t}
    J_t 
     = -\nu^2 \,\frac{2}{\pi} \int dk\, \frac{\cos\left(k \frac{t}{\tau}\right)}{(1 + k^2) \sqrt{2 +k^2}}\, e^{-\frac{|t|}{\tau} \sqrt{2 +k^2} },
\end{equation}
and its local nature is shown by the upper bound $|J_t| \leq \nu^2   \exp(-\sqrt{2} |t|/\tau )$ in terms of the correlation timescale $\tau$.


The perturbative corrections to the expectation are also decomposed into pure nonlinearities and shape corrections. The pure nonlinearities can be derived from the fluctuation theorem, $\langle \exp(\widetilde{x}) \rangle = 1$, but for consistency we also checked the corresponding diagrams in the SM. The shape corrections are evaluated as, up to $\mathcal{O}\left(\nu^2\right)$,
\begin{multline}
\langle x_{0, 0} \rangle - \langle \widetilde{x}_{0, 0} \rangle  \quad = \quad
\vcenter{\hbox{\includegraphics[page=7]{Figures.pdf}}}
\\ \quad + \quad
\vcenter{\hbox{\includegraphics[page=8]{Figures.pdf}}}
\quad + \quad
\vcenter{\hbox{\includegraphics[page=9]{Figures.pdf}}}
\quad \\[5pt]
+ \quad
\vcenter{\hbox{\includegraphics[page=10]{Figures.pdf}}}
\quad + \quad
\vcenter{\hbox{\includegraphics[page=11]{Figures.pdf}}}
\\[7pt]
=- \frac{\alpha}{2} \left[ J_{t} \,\overline{f^{t}} +K_{t, t'}\, \overline{f^{t} f^{t'}}
    - J_{t} \, \overline{f_0 \left(\overline{f^{t}} -f^{t}/\alpha \right)}   \right] .
\end{multline}

\acknowledgments
GC acknowledges LIMS Cognia visitor program.

\bibliographystyle{eplbib}
\bibliography{bibliography}

\clearpage
\includepdf[pages=-]{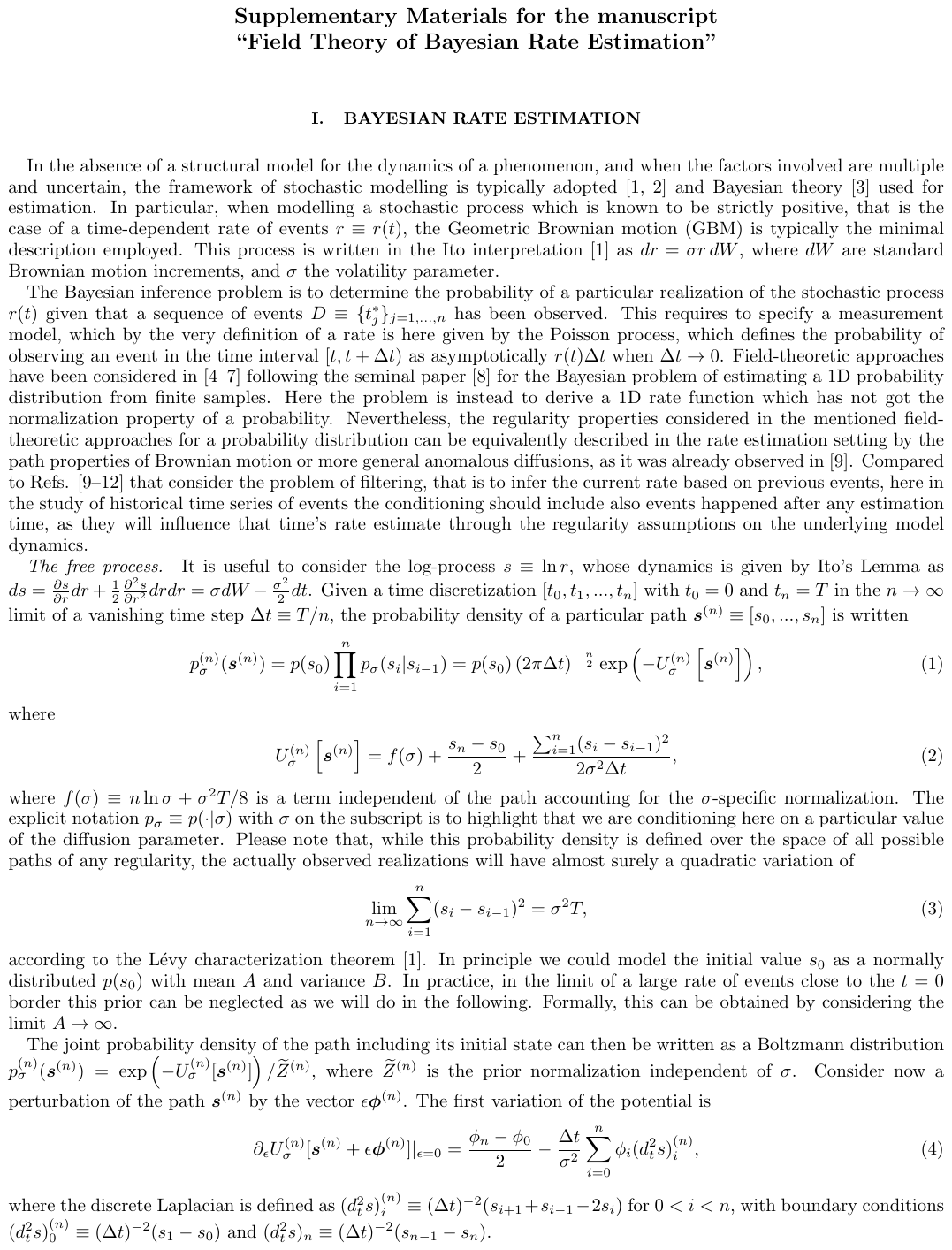}

\end{document}